\documentclass[12pt]{article}

 \ifx\pdfoutput\undefined
   \usepackage[dvips]{graphicx}
   \else
   \usepackage[pdftex]{graphicx}
   \fi
   \usepackage{epstopdf}

\newcommand{\be}{\begin{eqnarray}}
\newcommand{\ee}{\end{eqnarray}}
\newcommand{\sll}{\raise.15ex\hbox{$/$}\kern-.43em\hbox{$l$}}
\newcommand{\slp}{\raise.15ex\hbox{$/$}\kern-.43em\hbox{$p$}}
\newcommand{\slq}{\raise.15ex\hbox{$/$}\kern-.43em\hbox{$q$}}
\newcommand{\slk}{\raise.15ex\hbox{$/$}\kern-.43em\hbox{$k$}}
\newcommand{\slepsilon}{\raise.15ex\hbox{$/$}\kern-.53em\hbox{$\epsilon$}}

\begin{document}
\DeclareGraphicsExtensions{.pdf,.gif,.jpg}

\bibliographystyle{unsrt}
\footskip 1.0cm

\thispagestyle{empty}
\begin{flushright}
BCCUNY-HEP- 07-02
\end{flushright}
\vspace{0.1in}

\begin{center}{\Large \bf {Rapidity dependence of the photon to pion production ratio in high energy collisions}}\\

\vspace{1in}
{\large  Jamal Jalilian-Marian}\\

\vspace{.2in}
{\it Department of Natural Sciences, Baruch College, New York, NY 10010\\ }

\end{center}

\vspace*{25mm}

\begin{abstract}

\noindent We investigate rapidity dependence of the ratio of photon and pion production cross sections in high energy proton (deuteron) - nucleus collisions at RHIC and LHC. This ratio, and its rapidity dependence can be a sensitive probe of high density QCD (Color Glass Condensate) dynamics and shed further light on the role of saturation physics at RHIC and LHC.

\end{abstract}
\newpage

\section{Introduction}
The Color Glass Condensate (saturation) approach to QCD at high energy \cite{glr,nonlin,bk} has been quite successful in description of variety of high energy processes at RHIC \cite{rhic}. From energy, rapidity and centrality dependence of hadron multiplicities \cite{kn} to $p_t$ dependence of less inclusive observables such as hadron production in deuteron-gold collisions \cite{pheno,dhj}, there is growing evidence for importance and/or dominance of saturation physics at RHIC (for a review see \cite{ykjjm2} and references therein). Nevertheless, in order to further test saturation physics and probe the limits of its validity, it is highly desirable to consider more processes which can be measured at RHIC and/or will be measured at LHC. Here we focus on the ratio of photon to pion production cross sections at different rapidities and at fixed transverse momentum which can be measured at RHIC and which can be used to further probe the saturation dynamics and its applicability at different regions of RHIC kinematics.

Even though the Color Glass Condensate formalism has been very successful in description of the $p_t$ dependence of hadronic cross sections at RHIC \cite{dhj}, there are aspects of the approach which need further refinement. For example, in order to describe the pion production cross section at different rapidities, one needs a K factor in order to reproduce the overall magnitude of the cross section. This by itself is not a cause for concern since it is well known that a  LO QCD description of hadron production cross section receives large contributions from higher order corrections in perturbation theory. While these higher order corrections are known in pQCD, the Color Glass Condensate, at its current level of sophistication, is known only at the Leading Order in the coupling constant (for a calculation of running of the coupling constant see \cite{kwb}). What is essential is that the needed K factor is independent of $p_t$. Nevertheless, it is desirable to consider observables which are not very sensitive to the K factor.

In this brief note, we investigate the ratio of photon to pion production cross sections at different rapidities. The hope is that the K factors involved will more or less cancel so that this ratio will not be very sensitive to it. This may be motivated by the fact that at low $p_t$ (typically less than $2-3$ GeV) most of the produced photons (not coming from hadron decays) are due to fragmentation of a produced quark (or anti-quark) and are therefore similar to the hard processes involved in hadron production. The price to pay for this is that one needs to consider low $p_t$ photons in order to minimize contribution of direct photons. This will no doubt complicate the experimental efforts but it can in principle be done at RHIC and LHC.

After a brief review of the analytic forms of the pion \cite{dhj}  and photon \cite{fgjjm,bmds} production cross sections in the Color Glass Condensate formalism in terms of the dipole cross section (both processes have adjoint and/or fundamental dipole cross sections as the universal building blocks), we use a parameterization of the dipole cross section which has been used before to successfully describe the RHIC data on hadron production in deuteron-gold collisions at RHIC. We first apply our formalism to proton-proton collisions at RHIC and calculate the ratio of photon to pion production cross section at different rapidities and compare our result with the known NLO pQCD predictions \cite{nlo}. It is observed that at mid rapidity, our result is below that given by NLO QCD while at forward rapidities the two are in pretty good agreement. This can be understood as being due to dominance of gluons in mid rapidity which means one would need to include contributions due to fragmentation of gluons to photons. This is formally a higher order (in $\alpha_s$) effect but is numerically large. Furthermore, application of Color Glass Condensate formalism  (which is inherently a weak coupling approach) to proton-proton collisions at mid rapidity RHIC is rather questionable since at mid rapidity RHIC the saturation scale of a proton is rather small. We then investigate this ratio for deuteron-gold collisions at RHIC and proton-lead collisions at LHC. Our results are therefore most reliable at the very forward rapidities at RHIC and LHC where quark production dominates over gluon production. We comment on how one may include higher order corrections due to contribution of gluons.

\section{The photon and pion production cross sections}

Here we briefly review the photon and pion production cross sections in the hybrid approach \cite{dhj,fgjjm,adjjm,bms} to particle production in the Color Glass Condensate formalism. The target is described as a dense system of gluons which satisfy the JIMWLK evolution equations while the projectile is described as a collection of partons in pQCD which evolve via DGLAP evolution 
equations (for an alternative formulation which describes the projectile using CGC formalism, see \cite{bgv,ykam}). The photon production cross section is given by 
\be
{d\sigma^{p(d)\, A \rightarrow \gamma(p_t, y_{\gamma})\, X}\over 
d^2b_t\, d^2p_t\, dy_{\gamma}}\!\!\! &=& \!\!\!
{1\over (2\pi)^2} \sum_f  \int dx_q \,
[q_f(x_q,p_t^2) + \bar{q}_f(x_q,p_t^2)]\, {D_{\gamma/q}(z,p_t^2) \over z}
N_F (x_g,{p_t\over z},b_t)
\label{eq:cs_pho}
\ee
where $N_F$ is the probability for scattering of a fundamental (quark-anti-quark) dipole on the dense target and $D_{\gamma/q}$ is the LO pQCD quark-photpn fragmentation function. Depending on the projectile being a proton or deuteron, one sums over the appropriate quark and anti-quark distributions. Here, we ignore the possible nuclear modifications of the deuteron wavefunction since these are expected to be small in this kinematic \cite{hkn} and will mostly cancel in the ratio. Furthermore, the fraction of the incoming target momentum carried by the gluons is denoted $x_g$ and is given by 
$x_g = x_q\, e^{-2 y_{\gamma}}$ while $z = {p_t \over x_q\, \sqrt{s}}\, e^{y_{\gamma}}$ 
and the lower limit in the $x_q$ integration is $x_q^{min}={p_t\over \sqrt{s}} e^{y_{\gamma}}$. 

The hadron production cross section is given by 
\be
{d\sigma^{p(d) A \rightarrow h(p_t, y_h)\, X} \over d^2 b_t\, d^2 p_t\, d y_h } &=& 
{1 \over (2\pi)^2}
\int_{x_F}^{1} dx_p \, {x_p\over x_F} \Bigg[
f_{q/p} (x_p,p_t^2)~ N_F \left(x_g, {x_p\over x_F} p_t , b_t\right)~ D_{h/q} 
\left({x_F\over  x_p}, p_t^2\right) +  \nonumber \\
&&
f_{g/p} (x_p,p_t^2)~ N_A \left(x_g, {x_p\over x_F} p_t , b_t\right) ~ 
D_{h/g} \left({x_F\over x_p}, p_t^2\right)\Bigg]
\label{eq:cs_had}
\ee
where $x_F = {p_t \over \sqrt{s}}\, e^{y_h}$ is the Feynman $x$ of the produced hadron and 
$x_g  = x_p\, e^{-2 y_h}$ where $x_p$ is the fraction of the incoming hadron energy carried by
the incoming parton. We also note that ${1\over z} = {x_p \over x_F}$ which relates the 
variables in  eqs. (\ref{eq:cs_pho}) and (\ref{eq:cs_had}). To proceed further, we need to know 
the fundamental and adjoint dipole cross sections $N_F,\, N_A$. Here we will use a parameterization 
of the adjoint dipole cross section which was used in order to investigate saturation effects in hadron production in deuteron-gold collisions at RHIC. In this parameterization, the adjoint dipole scattering probability (impact parameter integrated cross section) is given by 
\be  
\label{eq:NA_param}
N_A(r_t,y_h) = 1-\exp\left[ - \frac{1}{4} [r_t^2 Q_s^2(y_h)]^
{\gamma(y_h,r_t)}\right]
\ee
where the anomalous dimension $\gamma$ is given by 
\be
\gamma(r_t,Y) &=& \gamma_s + \Delta\gamma(r,Y) \nonumber\\
\Delta\gamma &=& (1-\gamma_s)\,\frac{\log (1/r_t^2\,Q_s^2)}{\lambda\, Y
+\log(1/r_t^2\,Q_s^2) + d\sqrt{Y}}~.  \label{eq:gam_new} 
\ee 
The expression for the scattering probability for a fundamental dipole can be obtained from 
(\ref{eq:NA_param}) by a simple re-scaling of the saturation scale. 
Since this parameterization was used to fit the minimum bias deuteron-gold data at RHIC, we will 
consider only the minimum bias cross sections here as well (for details of this parameterization, 
we refer the reader to \cite{dhj}). We now use eqs. (\ref{eq:cs_pho}) and (\ref{eq:cs_had}) to calculate the single inclusive photon and single inclusive pion production cross sections in proton-proton and proton (deuteron)-nucleus collisions in kinematics relevant to the RHIC and LHC experiments. 

\begin{figure}[hbtp]
   \begin{center}
   \includegraphics[width=4in]{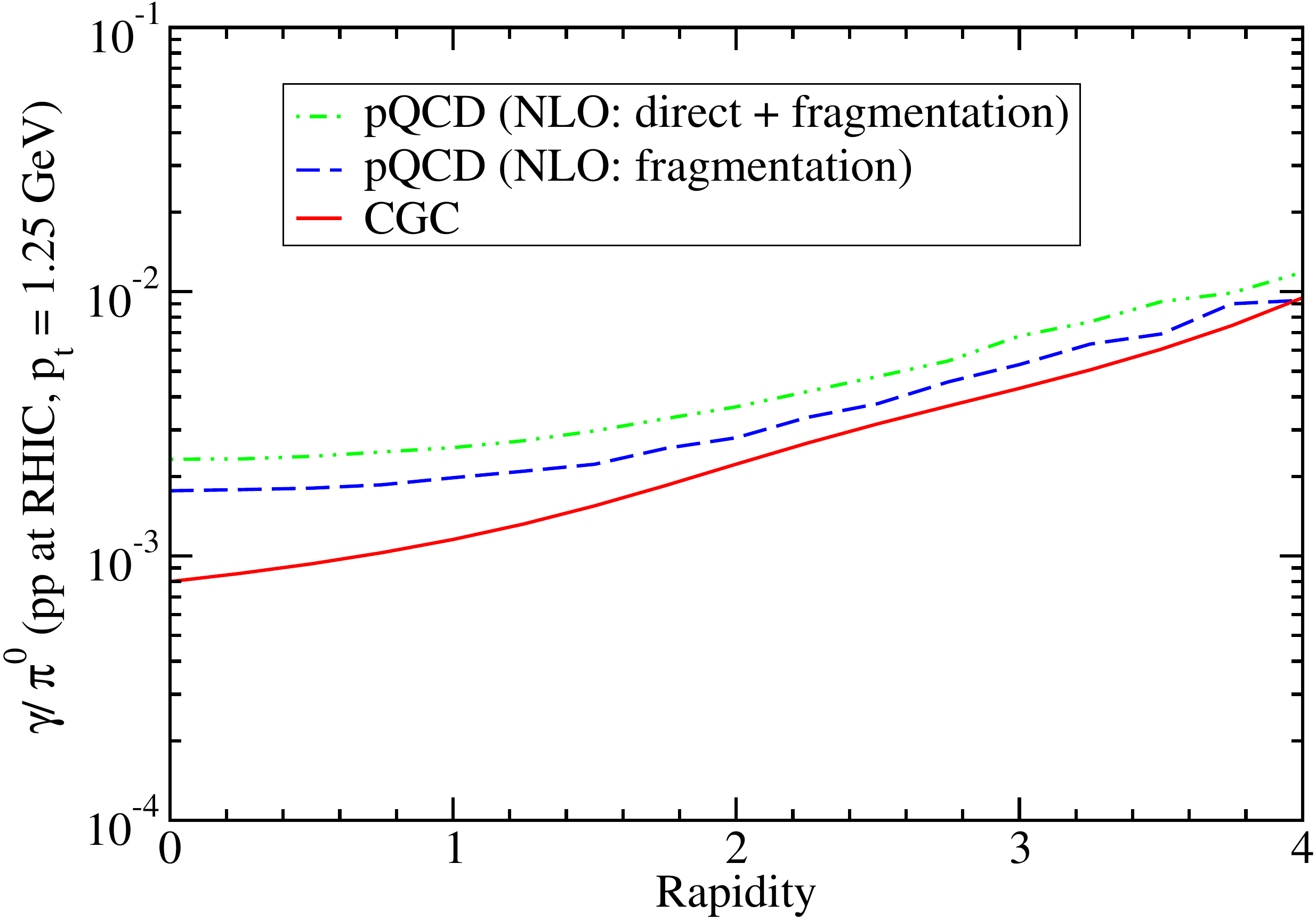}
   \caption{Ratio of fragmentation photon to neutral pion production cross sections in proton-proton collisions at RHIC.}
   \label{fig:pp_rhic}
   \end{center}
   \end{figure}

In Figure (\ref{fig:pp_rhic}) we show the ratio of photon to pion production cross sections in 
proton-proton collisions 
at RHIC. Our results, labeled CGC are shown against the NLO pQCD calculations, with and without
direct photons which contribute about $20-30\%$ at $p_t = 1.25$ GeV. 
As is seen, CGC results
are below NLO pQCD results by a factor of $\sim 2$ at mid rapidity and come close to pQCD results
only at the most forward rapidity. 
This is most likely due to the dominance of gluon contribution to photon production in mid rapidity. Even though this is nominally a higher order correction suppressed by $\alpha_s$, since low $p_t$ and mid rapidity kinematics probe small $x$ gluons in the incoming hadron wave function, this factor of $\alpha_s$ is easily compensated by the number of gluons in the wave function of the incoming hadron. Furthermore, the saturation scale of a proton probed in the mid rapidity region of RHIC is most likely not large enough to warrant a CGC type approach. Another possible problem is that the hybrid approach developed in is valid for asymmetric collisions, i.e. when a dilute projectile scatters on a dense target which is not the case in observables produced in mid rapidity in proton-proton collisions. Therefore, our results for pp collisions in mid rapidity region of RHIC have to be taken with a grain of salt and are shown here only for sake of comparison and completeness. 
     
\begin{figure}[htbp]
   \begin{center}
   \includegraphics[width=4in]{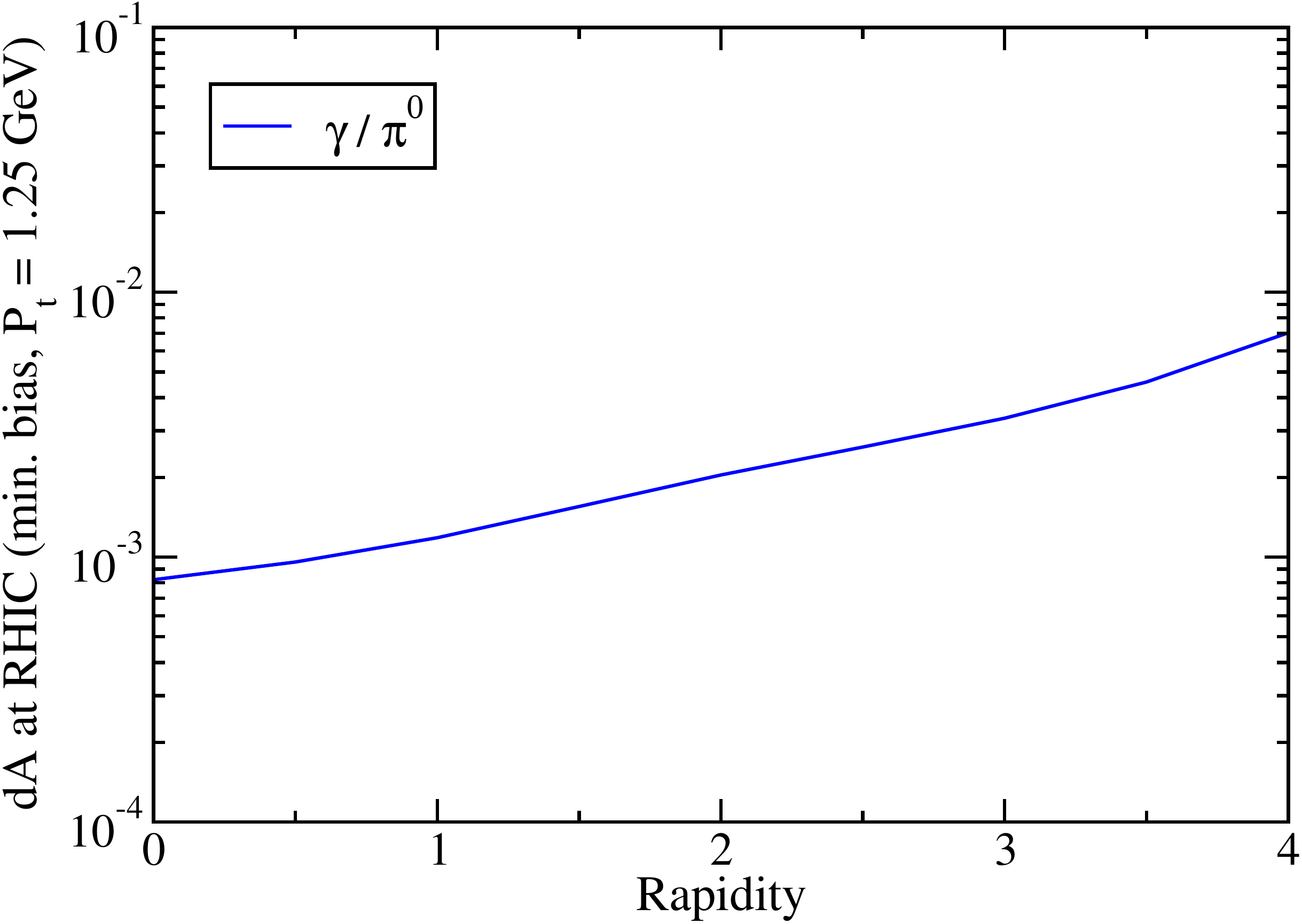}
   \caption{Ratio of fragmentation photon to neutral pion production cross sections in min. bias deuteron-gold collisions at RHIC.}
   \label{fig:dA_rhic}
   \end{center}
   \end{figure}

In Figure (\ref{fig:dA_rhic}) we show the ratio of fragmentation photon to neutral pion production cross sections in deuteron-gold collisions at RHIC and for minimum bias collisions. 
Again, a rise in the ratio is seen as one goes to more forward rapidities. This is mostly due to the diminishing contribution of gluons in the projectile hadron. In this case, the underlying hard scattering becomes identical for both fragmentation photon and pion production ({\it only $N_F$ contributes}) and the only difference between the two is due to the final state fragmentation functions (the average value
of $z$ is also somewhat different between photon and hadron fragmentation functions).
Having a nuclear rather than a proton target makes the saturation scale of the target larger and application of CGC formalism to this process is more robust. In this sense, it would even be better to consider the most central class of collisions since the saturation scale would be the largest, however, this is numerically quite complicated since it would require a Monte-Carlo simulation of centrality classes which is beyond the scope of this work.     
 
\begin{figure}[htbp]
   \begin{center}
   \includegraphics[width=4in]{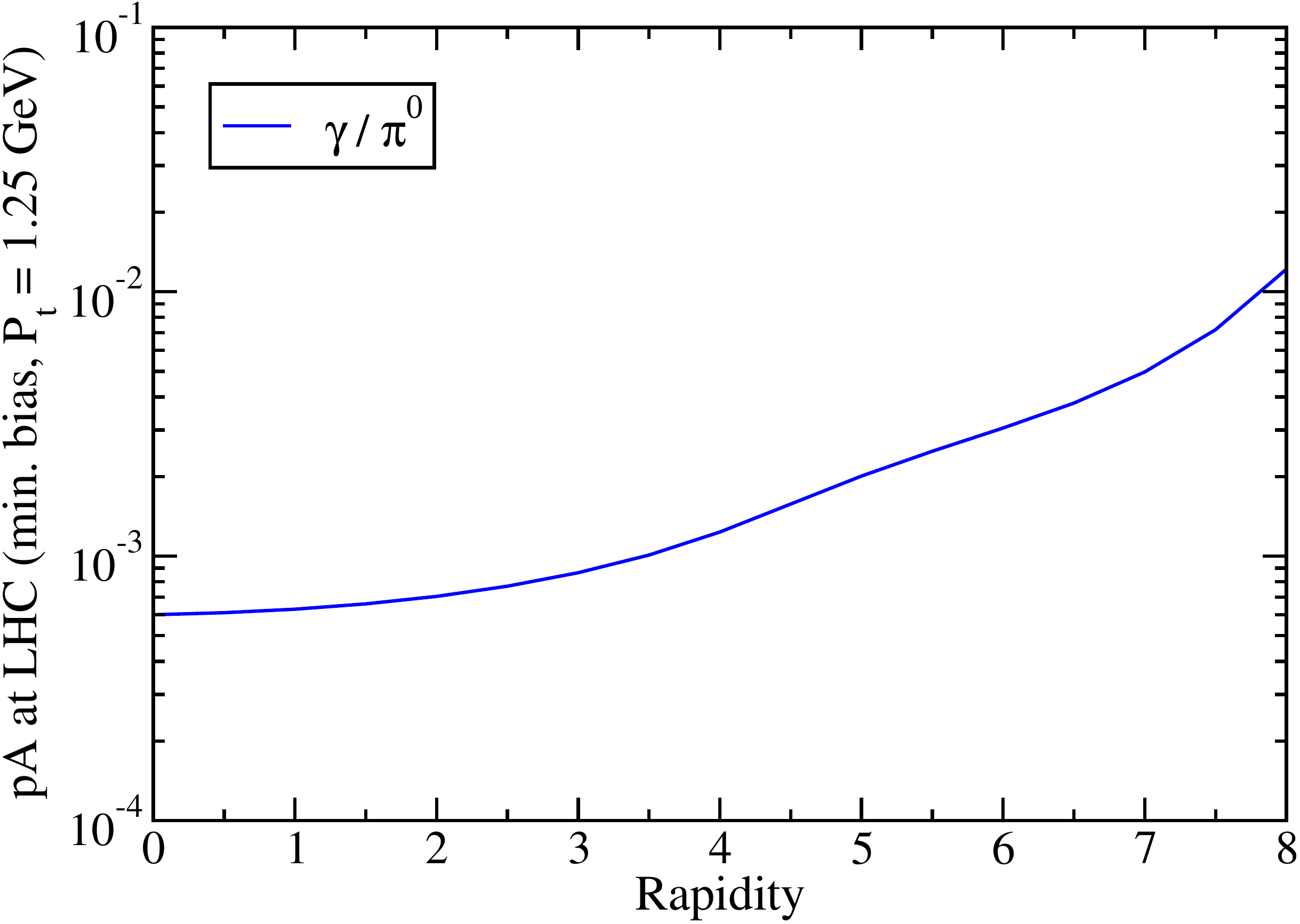}
   \caption{Ratio of fragmentation photon to neutral pion production cross sections  in min. bias proton-lead collisions at LHC.}
   \label{fig:pA_lhc}
   \end{center}
   \end{figure}

 In Figure (\ref{fig:pA_lhc}) we show the same ratio for the upcoming proton-lead collisions at LHC center of mass energy of $\sqrt{s} = 8.8$ GeV (we take the value for minimum bias saturation scale to be the same for a lead nucleus as for a gold nucleus. This approximation does not make a big numerical difference.). This ratio shows the same characteristics as before where a rather sharp rise is seen. We note that this ratio is smaller in mid rapidity LHC as compared to RHIC due to larger gluon production cross section at LHC. One should keep in mind that contribution of gluons to fragmentation photon may be quite large in mid rapidity which would increase this ratio in mid and not so forward rapidities (at LHC, quark production dominates gluon production only after $5-6$ units of rapidity away from mid rapidity). Therefore, calculation of higher order corrections to photon production in mid to not so forward rapidity seem to be essential to a precise computation of this ratio. Therefore we expect our results to be most accurate in the forward rapidity region, in both RHIC and LHC.

A precise measurement of ratio of photon to pion production cross sections at RHIC and LHC would therefore help establish the extent to which a CGC based approach may be applicable in different kinematics regions of RHIC and LHC and shed light on the importance of higher order corrections to the Color Glass Condensate formalism. While this is an experimentally challenging task, it is certainly possible to do and one hopes that the results presented here will provide the motivation to perform these measurements at RHIC and LHC.

\vspace{0.2in}
\leftline{\bf Acknowledgments} 

\noindent We thank F. Gelis for helpful discussions and the use of his Fourier transform code, W. Voglesang for useful discussions and providing us with NLO pQCD results and acknowledge helpful correspondence with S. Kumano regarding HKN parton distributions.

\vspace{0.2in}
\leftline{\bf References}

\renewenvironment{thebibliography}[1]
        {\begin{list}{[$\,$\arabic{enumi}$\,$]}  
        {\usecounter{enumi}\setlength{\parsep}{0pt}
         \setlength{\itemsep}{0pt}  \renewcommand{\baselinestretch}{1.2}
         \settowidth
        {\labelwidth}{#1 ~ ~}\sloppy}}{\end{list}}


\begin{thebibliography}{99}

\small

\bibitem{glr}
L. Gribov, E. Levin and M.Ryskin, {\it Nucl. Phys. }{\bf B188}, 555 (1981);
A.H. Mueller and J. Qiu, {\it Nucl. Phys. }{\bf B268}, 427 (1986).

\bibitem{nonlin}
L.~D.~McLerran and R.~Venugopalan,
Phys.\ Rev.\ D {\bf 49}, 2233 (1994),  
Phys.\ Rev.\ D {\bf 49}, 3352 (1994); 
A.~Ayala, J.~Jalilian-Marian, L.~D.~McLerran and R.~Venugopalan,
Phys.\ Rev.\ D {\bf 52}, 2935 (1995),
Phys.\ Rev.\ D {\bf 53}, 458 (1996); 
J. Jalilian-Marian, A. Kovner, L. McLerran and H. Weigert, 
{\it Phys. Rev. }{\bf D55}, 5414 (1997); 
J.~Jalilian-Marian, A.~Kovner, A.~Leonidov and H.~Weigert,
Nucl.\ Phys.\ B {\bf 504}, 415 (1997), 
Phys.\ Rev.\ D {\bf 59}, 014014 (1999),
Phys.\ Rev.\ D {\bf 59}, 034007 (1999),
[Erratum-ibid.\ D {\bf 59}, 099903 (1999)];
J.~Jalilian-Marian, A.~Kovner and H.~Weigert,
Phys.\ Rev.\ D {\bf 59}, 014015 (1999);
A.~Kovner, J.~G.~Milhano and H.~Weigert,
Phys.\ Rev.\ D {\bf 62}, 114005 (2000);
A.~Kovner and J.~G.~Milhano,
Phys.\ Rev.\ D {\bf 61}, 014012 (2000);
E.~Iancu, A.~Leonidov and L.~D.~McLerran,
Nucl.\ Phys.\ A {\bf 692}, 583 (2001),
hep-ph/0202270, 
Phys.\ Lett.\ B {\bf 510}, 133 (2001); 
E.~Iancu and L.~D.~McLerran,
Phys.\ Lett.\ B {\bf 510}, 145 (2001);
E.~Ferreiro, E.~Iancu, A.~Leonidov and L.~McLerran,
Nucl.\ Phys.\ A {\bf 703}, 489 (2002).

\bibitem{bk}
I.~Balitsky, Nucl.\ Phys.\ B {\bf 463}, 99 (1996);  
Y.~V.~Kovchegov, Phys.\ Rev.\ D {\bf 60}, 034008 (1999),
Phys.\ Rev.\ D {\bf 61}, 074018 (2000).

\bibitem{rhic}
I.~Arsene {\it et al.}  [BRAHMS Collaboration], nucl-ex/0410020; 
B.~B.~Back {\it et al.}, [PHOBOS Collaboration], nucl-ex/0410022; 
K.~Adcox {\it et al.}  [PHENIX Collaboration], nucl-ex/0410003; 
J.~Adams  [STAR Collaboration], nucl-ex/0501009.

\bibitem{kn}
  D.~Kharzeev and M.~Nardi,
  Phys.\ Lett.\  B {\bf 507}, 121 (2001)
  [arXiv:nucl-th/0012025].

\bibitem{pheno}
D.~Kharzeev, Y.~V.~Kovchegov and K.~Tuchin,
Phys.\ Rev.\ D {\bf 68}, 094013 (2003), 
Phys.\ Lett.\ B {\bf 599}, 23 (2004); 
J.~Jalilian-Marian,
  Nucl.\ Phys.\  A {\bf 748}, 664 (2005)
  [arXiv:nucl-th/0402080]; 
J.~L.~Albacete, N.~Armesto, A.~Kovner, C.~A.~Salgado and U.~A.~Wiedemann,
Phys.\ Rev.\ Lett.\  {\bf 92}, 082001 (2004); 
J.~Jalilian-Marian, Y.~Nara and R.~Venugopalan,
Phys.\ Lett.\ B {\bf 577}, 54 (2003). 

\bibitem{dhj}
  A.~Dumitru, A.~Hayashigaki and J.~Jalilian-Marian,
  Nucl.\ Phys.\  A {\bf 770}, 57 (2006)
  [arXiv:hep-ph/0512129], 
Nucl.\ Phys.\  A {\bf 765}, 464 (2006)
  [arXiv:hep-ph/0506308].

\bibitem{ykjjm2}
  J.~Jalilian-Marian and Y.~V.~Kovchegov,
  Prog.\ Part.\ Nucl.\ Phys.\  {\bf 56}, 104 (2006)
  [arXiv:hep-ph/0505052].

\bibitem{kwb}
  Y.~V.~Kovchegov and H.~Weigert,
  Nucl.\ Phys.\  A {\bf 784}, 188 (2007)
  [arXiv:hep-ph/0609090]; 
 I.~Balitsky,
  Phys.\ Rev.\  D {\bf 75}, 014001 (2007)
  [arXiv:hep-ph/0609105].
 
\bibitem{fgjjm}
  F.~Gelis and J.~Jalilian-Marian,
    arXiv:hep-ph/0609066, 
 Phys.\ Rev.\  D {\bf 67}, 074019 (2003)
  [arXiv:hep-ph/0211363], 
 Phys.\ Rev.\  D {\bf 66}, 094014 (2002)
  [arXiv:hep-ph/0208141], 
 Phys.\ Rev.\  D {\bf 66}, 014021 (2002)
  [arXiv:hep-ph/0205037].
 
\bibitem{bmds}
  R.~Baier, A.~H.~Mueller and D.~Schiff,
  Nucl.\ Phys.\  A {\bf 741}, 358 (2004)
  [arXiv:hep-ph/0403201].
 
\bibitem{nlo}
  P.~Aurenche, P.~Chiappetta, M.~Fontannaz, J.~P.~Guillet and E.~Pilon,
  Nucl.\ Phys.\  B {\bf 399}, 34 (1993); 
P.~Aurenche, R.~Baier, M.~Fontannaz and D.~Schiff,
Nucl.\ Phys.\ B {\bf 297}, 661 (1988).

 \bibitem{adjjm}
A.~Dumitru and J.~Jalilian-Marian,
Phys.\ Lett.\ B {\bf 547}, 15 (2002), 
Phys.\ Rev.\ Lett.\  {\bf 89}, 022301 (2002).

\bibitem{bms}
  R.~Baier, Y.~Mehtar-Tani and D.~Schiff,
  Nucl.\ Phys.\  A {\bf 764}, 515 (2006)
  [arXiv:hep-ph/0508026].

\bibitem{bgv}
F.~Gelis and R.~Venugopalan,
Phys.\ Rev.\ D {\bf 69}, 014019 (2004);
J.~P.~Blaizot, F.~Gelis and R.~Venugopalan,
Nucl.\ Phys.\ A {\bf 743}, 13 (2004), 
Nucl.\ Phys.\ A {\bf 743}, 57 (2004); 
  F.~Gelis and Y.~Mehtar-Tani,
  Phys.\ Rev.\  D {\bf 73}, 034019 (2006)
  [arXiv:hep-ph/0512079].

\bibitem{ykam}
  Y.~V.~Kovchegov and A.~H.~Mueller,
  Nucl.\ Phys.\  B {\bf 529}, 451 (1998)
  [arXiv:hep-ph/9802440].
  
\bibitem{hkn}
  M.~Hirai, S.~Kumano and T.~H.~Nagai,
  Phys.\ Rev.\  C {\bf 70}, 044905 (2004)
  [arXiv:hep-ph/0404093].


\end{thebibliography}
\end{document}